   \documentclass[preprint,showpacs,preprintnumbers,amsmath,amssymb,superscriptaddress]{revtex4}
\usepackage{graphicx} % Include figure files
\usepackage{dcolumn}  % Align table columns on decimal point
\usepackage{bm}       % bold math

\begin{document}

\preprint{}

\title{Coulomb effects in cold fission from reactions\ $^{233}$U($n_{\rm{th}}$,f) and $^{235}$U($n_{\rm{th}}$,f)}

\author{M. Montoya}
\email{mmontoya@ipen.gob.pe}

\affiliation{Instituto Peruano de Energ\'ia Nuclear, Canad\'a 1470, San Borja, Lima, Per\'u.}
\affiliation{Facultad de Ciencias, Universidad Nacional de Ingenier\'ia, Av. T\'upac Amaru 210, R\'imac, Lima, Per\'u.}

\date{\today} % It is always \today, today,
%\date{29 de Setiembre de 2008}

\begin{abstract}
 The Coulomb effect hypothesis in cold fission, formerly used to interpret fluctuations in the curve of maximal total kinetic energy as a function of fragments mass, in reactions $^{233}$U($n_{\rm{th}}$,f), $^{235}$U($n_{\rm{th}}$,f) and $^{239}$Pu($n_{\rm{th}}$,f), is confirmed by the preference for more asymmetrical charge splits observed in cold fragmentations. Several experimental results on reactions $^{233}$U($n_{\rm{th}}$,f) and $^{235}$U($n_{\rm{th}}$,f) show that, for two isobaric fragmentations with similar $Q$-values, the greater is the value of the kinetic energy of the fragments, the greater will be the probability of more asymmetric charge split.
\end{abstract}

\pacs{24.75.+i,25.85.-w,21.10.sf,21.10.Gv,25.85.Ec,21.10.Ft}

 \maketitle

\section{Introduction}
\begin{figure}%[!h]
\centering
\includegraphics[width=8cm]{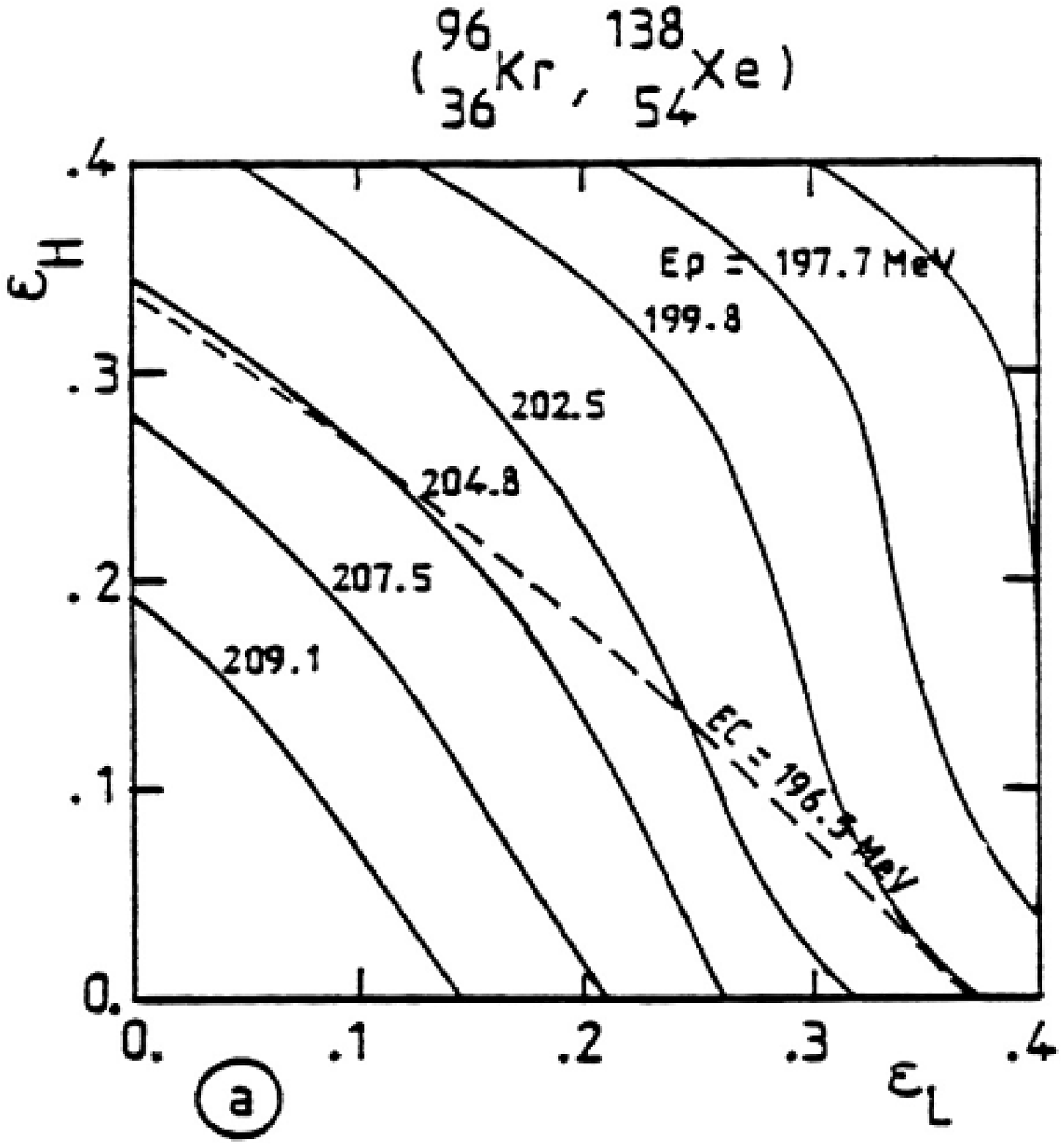}
\caption{ Potential energy ($P$, solid lines) equal to Coulomb interaction energy ($C$) plus total deformation energy ($D$), as a function of Nilsson´s deformation parameters of light (${\varepsilon }_{\rm{L}}$) and heavy (${\varepsilon }_{\rm{H}}$) fragments for ${}^{96}_{36}$Kr and $\ {}^{138}_{54}$Xe pair at scission point. The maximal Coulomb interaction energy obeying the balance energy relation $P\le Q$, $C_{\rm{max}}$, is 196.5 MeV. The minimal total deformation energy, $D_{\rm{min}}$, is around 3 MeV. }
\label{fig1}
\end{figure}

\begin{figure}%[!h]
\centering
\includegraphics[width=8cm]{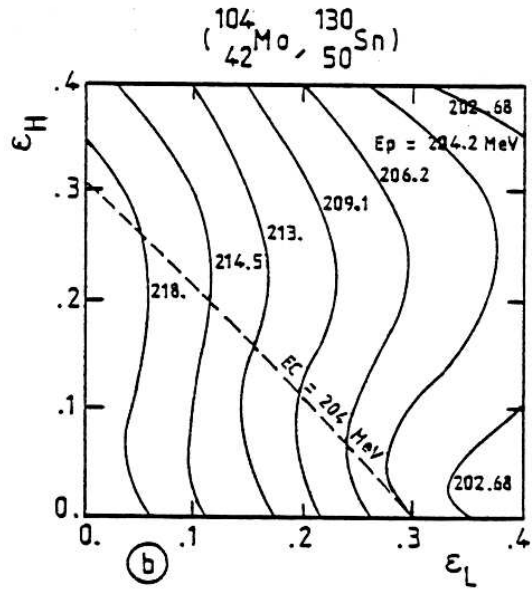}
\caption{ Similar to Fig.~\ref{fig2}, but corresponding to for ${}^{104}_{42}$Mo and $\ {}^{130}_{50}$Sn pair. $C_{\rm{max}}$ is equal to the corresponding $Q$-value. Taken from Ref. 5.}
\label{fig2}
\end{figure}

Among the most studied properties of nuclear fission of actinides are the distributions of mass and kinetic energy associated to complementary fragments \cite{1}. Pleasonton found that the highest total kinetic energy is around 190 MeV \cite{2}. However, those distributions are disturbed by neutron emission. In order to describe one of the consequences of neutron emission, let us suppose that a nucleus with proton number $Z_{\rm{f}}$ and mass number $A_{\rm{f}}$ splits into complementary light (L) and heavy (H) fragments corresponding to primary mass numbers $A_{\rm{L}}$ and $A_{\rm{H}}$, and proton numbers $Z_{\rm{L}}$ and $Z_{\rm{H}}$, having kinetic energies $E_{\rm{L}}$ and $E_{\rm{H}}$, respectively. After neutron emission, those fragments will end with mass numbers
\begin{equation*}
 m_{\rm{L}}=A_{\rm{L}}-n_{\rm{L}}
\end{equation*}
and
\begin{equation*}
 m_{\rm{H}}=A_{\rm{H}}-n_{\rm{H}}
\end{equation*}
where $n_{\rm{L}}$ and $n_{\rm{H}}$  are the numbers of neutrons emitted by the light and heavy fragments, respectively. The corresponding final kinetic energies associated to those fragments will be
\begin{equation*}
e_{\rm{L}} \cong E_{\rm{L}} \left(1- \dfrac{n_{\rm{L}}}{A_{\rm{L}}} \right)
\end{equation*}
and
\begin{equation*}
e_{\rm{H}} \cong E_{\rm{H}} \left(1- \dfrac{n_{\rm{H}}}{A_{\rm{H}}} \right)
\end{equation*}
respectively. At the Lohengrin spectrometer at the High Flux Reactor (HFR) of the Laue-Langevin Institute (ILL) in Grenoble, only light fragments are detected, in which case subscripts are omitted. Thus, for a given primary fragment mass number ($A$), the kinetic energy ($E$) distribution is characterized by the corresponding average ($\overline{E}(A)$) and standard deviation ($\sigma_{E}(A)$). In the case of reaction $^{235}$U($n_{\rm{th}}$,f), R. Brissot {\it et al.} \cite{3} showed that neutron emission from fragments, occurred before they reach the detector, generates a peak around $m=110$ on the $\sigma_{e}(m)$ curve, non-existent on the $\overline{E}(A)$ curve. In order to avoid neutron emission effects, the cold fission region, corresponding to highest values of kinetic energy, was studied by C. Signarbieux {\it et al.} \cite{4,5}. They choose high kinetic energy windows associated to fragments from reactions $^{233}$U($n_{\rm{th}}$,f), $^{235}$U($n_{\rm{th}}$,f) and $^{239}$Pu($n_{\rm{th}}$,f), respectively. The number of events analysed were 1.5$\times$10$^{6}$, 3$\times$10$^{6}$ and 3.2$\times$10$^{6}$, respectively. These authors used the difference of time of flight technique with solid detectors to measure the fragment kinetic energy. The experiment was conducted at the HFR of ILL. They succeeded to separate neighboring masses, which was the evidence that no neutron emission occurred, which permitted them to measure the maximal total kinetic energy as a function of primary fragment mass, $K_{\rm{max}}(A_{\rm{L}})$. \\\\
In order to reproduce those measured $K_{\rm{max}}(A_{\rm{L}})$ values, a scission point model is used. One assumes that the pre-scission kinetic energy is zero. Thus, the maximal total kinetic energy is a result of Coulomb repulsion between complementary fragments which begins at the scission point having the most compact configuration. \\\\
In general, the maximal total kinetic energy curve is lower that the maximal $Q$-value ($Q_{\rm{max}}$) curve. For instance, in the reaction $^{233}$U($n_{\rm{th}}$,f), the maximal total kinetic energy of the fragment pair ($^{96}$Kr, $^{138}$Xe) is 3 MeV lower than the corresponding $Q$-value, which is reproduced by a configuration composed by the heavy fragment $^{138}$Xe in its spherical shape and the light fragment $^{96}$Kr in a deformed state with ellipsoidal parameter $\epsilon = 0,4$. However, the maximal total kinetic energy of the fragment pair ($^{104}$Mo, $^{130}$Sn) reaches the corresponding $Q$-value, which is reproduced by a scission configuration conformed by the heavy fragment $^{130}$Sn in its spherical ground state, and the fragment $^{104}$Mo in its deformed ground state with an ellipsoidal parameter $\epsilon = 0,3$ \cite{5, 6, 7}. See Figs. \ref{fig1} and \ref{fig2}.\\\\
Calculations of Coulomb interaction energy ($C$) shows that, among neighboring fragment masses corresponding to similar $Q$-values, the lower is the light fragment charge ($Z_{\textrm L}$), the higher will be maximal Coulomb interaction energy ($C_{\textrm{max}})$, and, consequently, the higher will be the corresponding $K_{\rm{max}}$ value. This named Coulomb effect reproduces the observed fluctuations in experimental $K_{\rm{max}}(A_{\rm L})$ curves, with a period of 5 units of fragment mass, which is the average of the period of change in the fragment even charge that maximizes the $Q$-value \cite{5, 6, 7}.\\\\
In this paper, experimental results on charge yield in cold fragmentations will be interpreted as a confirmation of the Coulomb effect hypothesis.\\\\
\section{Coulomb effects on isobaric charge splits}
\begin{figure}%[!ht]
\centering
\includegraphics[width=8cm]{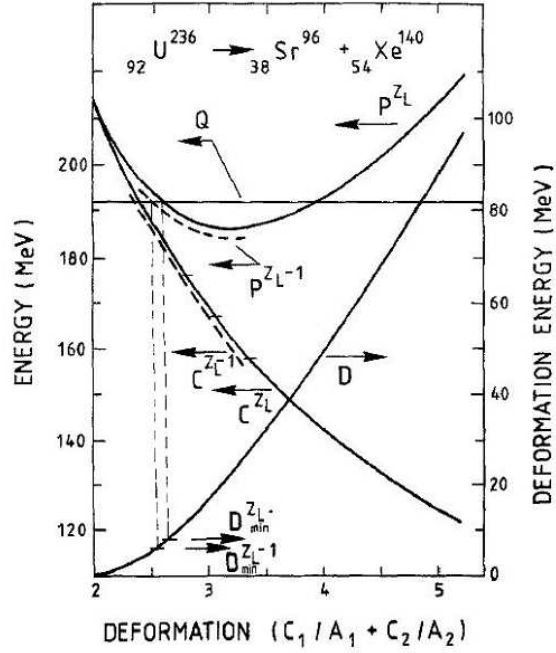}
\caption{Solid lines represent total deformation energy ($D$), Coulomb interaction energy ($C$) and potential energy  ($P=D+C$) curve of scission configurations as a function of deformation, defined as the sum of ratios of major to minor semi-axes of the fragments limited by the total available energy ($Q$) for the fragmentation ${}^{96}_{38}{{\rm{Kr}}}/{}^{140}_{54}{{\rm{Xe}}}$ (from Ref. 8). Dashed lines represent similar curves corresponding to the neighboring more asymmetrical charge split ($38-1)/(54+1$) having the same $Q$-value, for which one will obtain a maximal total kinetic energy higher than the one corresponding to the first fragmentation (from Ref. 7).}
\label{fig3}
\end{figure}

\begin{figure}%[!ht]
\centering
\includegraphics[width=8cm]{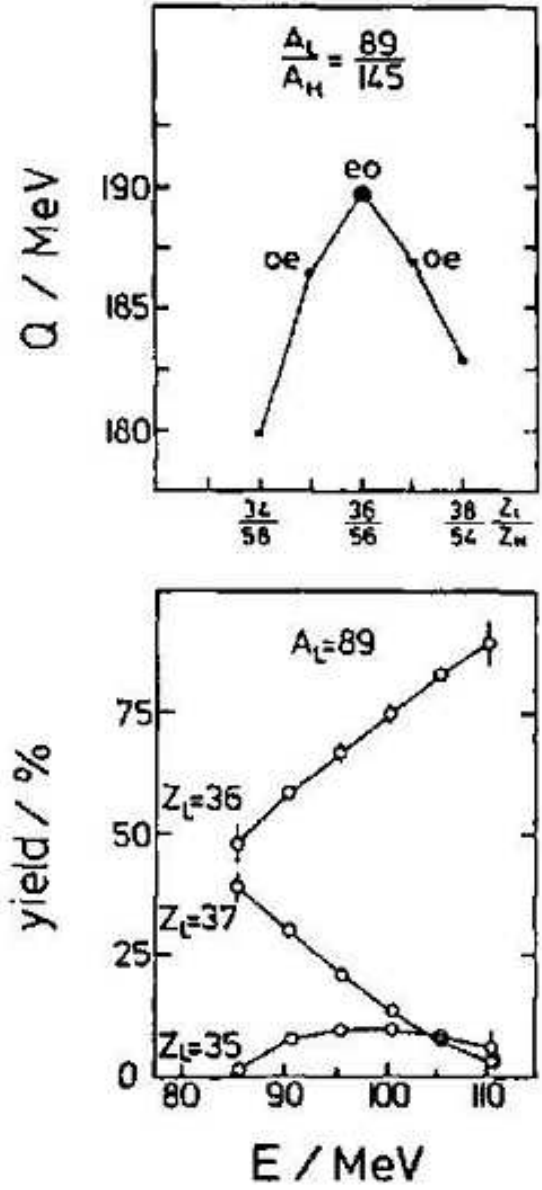}
\caption{Experimental yields (below) and calculated $Q$-values (above) for the odd-odd mass fragmentation $A_{\rm{L}}$/$A_{\rm{H}}{\rm =\ 89/145}$. The $Q$-values are calculated from the mass tables of Wapstra {\it et al.} [10]. Although charges $Z_{\rm{L}}=\ 35$ and $Z_{\rm{L}}=\ 37$ have similar $Q$-values, at $E_{\rm{L}}\ =\ 110.55$ MeV the higher probability corresponds to the lower light fragment charge.}
\label{fig4}
\end{figure}

\begin{figure}%[!ht]
 \centering
 \includegraphics[width=8cm]{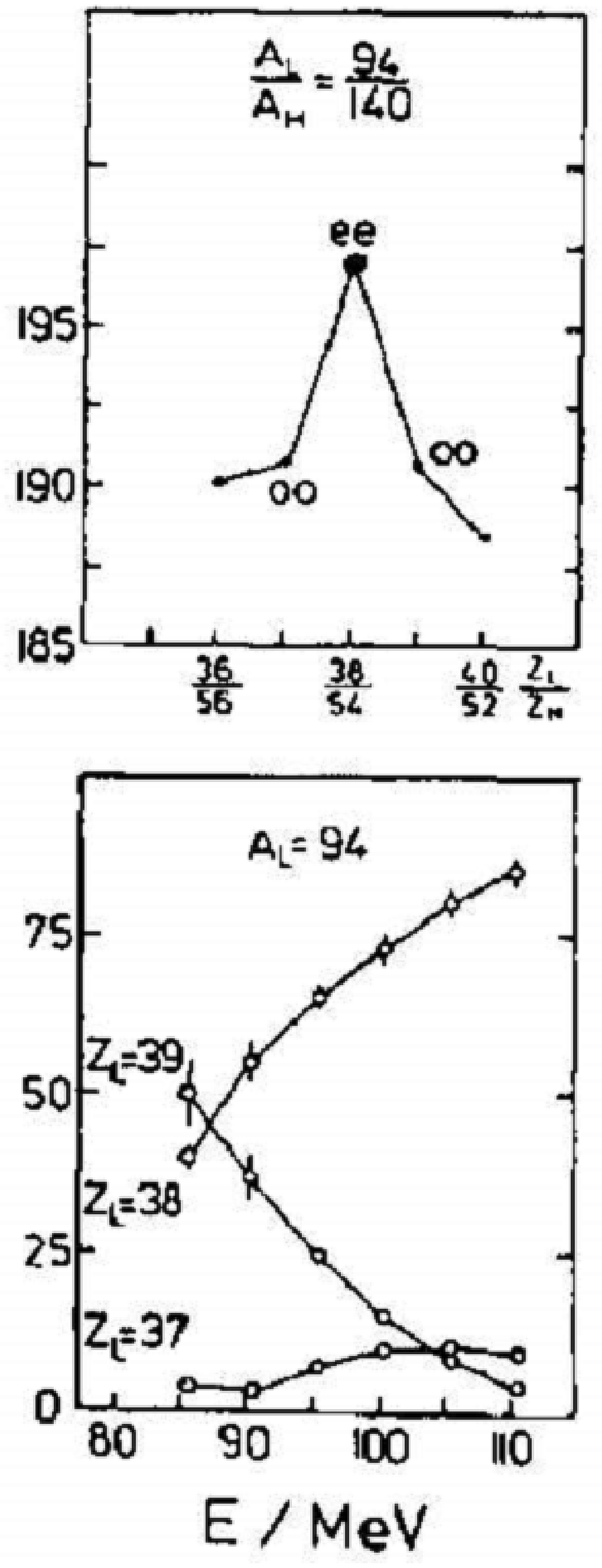}
 \caption{Similar to Fig.4 for $A_{\rm{L}}$/$A_{\rm{H}}{\rm =\ 94/140}$. Although charges $Z_{\rm{L}}=\ 37$ and $Z_{\rm{L}}=\ 39$, have similar $Q$-values, at $E_{\rm{L}}\ =\ 110.55$ MeV, the higher probability corresponds to the lower light fragment charge.}
 \label{fig5}
 \end{figure}

\begin{figure}%[!ht]
 \centering
 \includegraphics[width=8cm]{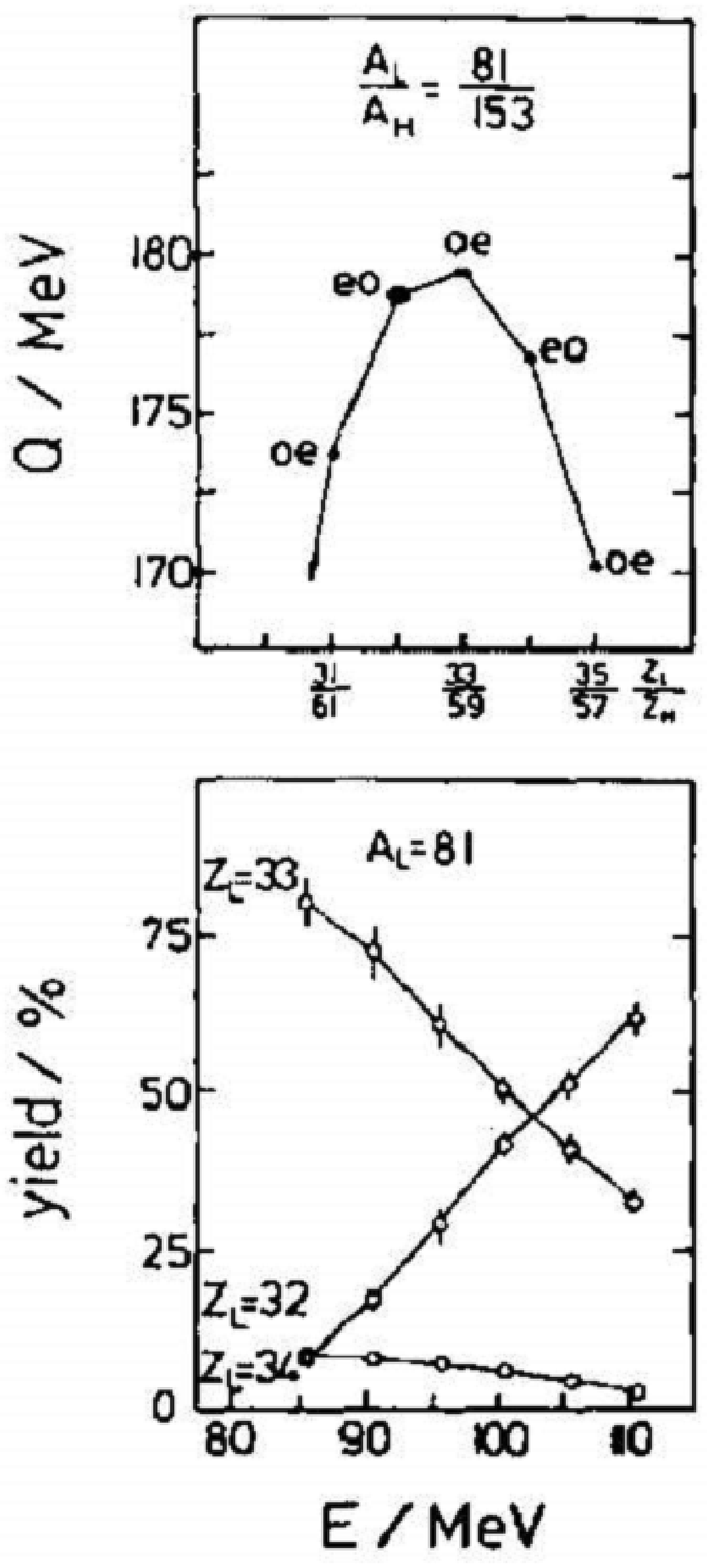}
 \caption{Similar to Fig. 4 for $A_{\rm{L}}$/$A_{\rm{H}}{\rm =\ 81/153}$. Although charges $Z_{\rm{L}}=\ 32$ and $Z_{\rm{L}}=\ 33$, have similar $Q$-values, at $E_{\rm{L}}\ =\ 110.55$ MeV, the higher probability corresponds to the lower light fragment charge.}
 \label{fig6}
 \end{figure}

\begin{figure}%[!ht]
 \centering
 \includegraphics[width=8cm]{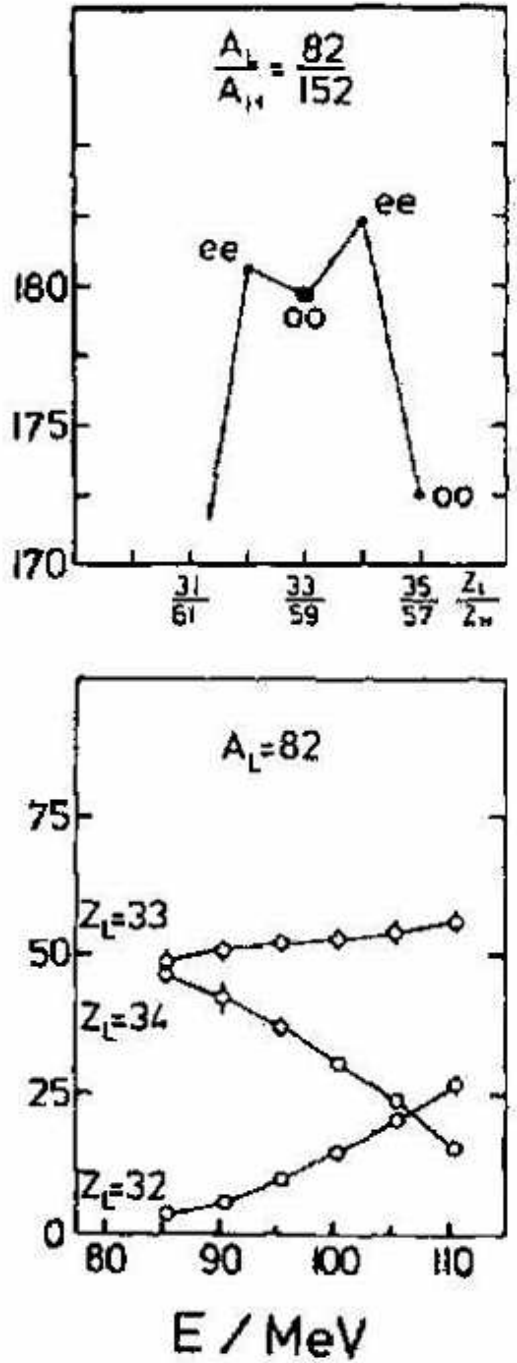}
 \caption{Similar to Fig. 4 for $A_{\rm{L}}$/$A_{\rm{H}}{\rm =\ 82/152}$. Although charges $Z_{\rm{L}}=\ 32$ and $Z_{\rm{L}}=\ 34$, although the higher $Q$-value correspond to the higher charge, at $E_{\rm{L}}\ =\ 110.55$ MeV, the higher probability corresponds to the lower light fragment charge. Moreover, the yield of the lower charge increases with higher slope than the other charges do. }
 \label{fig7}
 \end{figure}

\begin{figure}%[!ht]
 \centering
 \includegraphics[width=7cm]{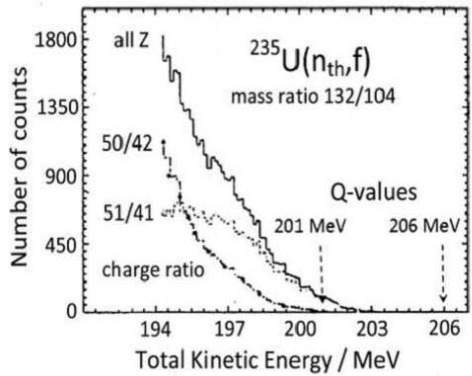}
 \caption{High energy $K$ tail for charge splits $50/42$ and $51/41$, corresponding to the mass fragmentation $132/104$ in $^{235}$U($n_{\rm{th}}$,f). Although the lower light fragment charge corresponds to lower $Q$-value, it has higher maximal total kinetic energy.}
 \label{fig8}
 \end{figure}

Let us assume that the composed nucleus characterized by mass  $A_{\rm{f}}$ and charge $Z_{\rm{f}}$ splits into two fragments, the light of which has a mass number $A_{\rm{L}}$ and a proton number $Z_{\rm{L}}$. In a scission point model, the potential energy ($P$) of a scission configuration is given
\begin{equation*}
P^{Z_{\rm{L}}}\left({\mathcal D}\right)=D^{Z_{\rm{L}}}\left({\mathcal D}\right)+C^{Z_{\rm{L}}}\left({\mathcal D}\right)
\end{equation*}
where $D$ is the total deformation energy of fragments, $C$ is the Coulomb interaction energy between complementary fragments, and ${\mathcal D}$ represents the deformed configuration shape. The scission configuration shapes are limited by the relation $P\le Q$.\\\\
The Coulomb interaction energy between two complementary hypothetical spherical fragments at scission configuration is given by
\begin{equation*}
C^{Z_{\rm{L}}}_{\rm{sph}}\ =\ \frac{1}{4\pi {\varepsilon }_0}·\frac{Z_{\rm{L}}(Z_{\rm{f}}-Z_{\rm{L}})e^2}{R_{\rm{L}}+R_{\rm{H}}+d}
\end{equation*}
where ${\varepsilon }_0$ is the electrical permittivity, $e$ is the electron charge, $R_{\rm{L}}$ and $R_{\rm{H}}$ are the radio of light and heavy fragment, respectively, and $d$ is the distance between surfaces of fragments (in this paper it is assumed that $d=2$ fm). The nucleus radio for each fragment is given by the relation $R=1.24A^{1/3}$ fm. Then, one can show that 
\begin{equation*}
\Delta C_{\rm{sph}}\left(Z_{\rm{L}},Z_{\rm{L}}-1\right)=C^{Z_{\rm{L}}}_{\rm{sph}}-C^{Z_{\rm{L}}-1}_{{\rm{sph}}}=\frac{(Z_{\rm{f}}-2Z_{\rm{L}}+1)}{Z_{\rm{L}}(Z_{\rm{f}}-Z_{\rm{L}})}C^{Z_{\rm{L}}}_{{\rm{sph}}}.
\end{equation*} 
In order to see how this value change with the charge fragmentation asymmetry, let us take two cases of charge splits from fission of nucleus $^{236}$U which has $Z_{\rm{F}}=92$. The first case corresponding to $Z_{\rm{L}}=46$ for which the relative variation $\Delta C_{\rm{sph}}$ produced by changing to $Z_{\rm{L}}-1=45$ will be nearly zero; and the second case, a much more asymmetric charge split, corresponding to $Z_L=30$, for which the variation $\Delta C_{\rm{sph}}$ produced by changing to $Z_{\rm{L}}-1=29$ will be approximately 3.5 MeV. As one can see, the Coulomb effect increases with asymmetry of charge split.\\\\
In general, the Coulomb interaction energy between spherical fragments is higher than the $Q-$value. Therefore, in a scission configuration, fragments must be deformed. Let us assume that, for isobaric split $A_{\rm{L}}$/$A_{\rm{H}}$, $C^{Z_{\rm{L}}}({\mathcal D})$ is the interaction Coulomb energy between complementary fragments corresponding to light charge $Z_{\rm{L}}$ and scission configuration shape ${\mathcal D}$, with fragments nearly spherical. If one takes two isobaric splits with light fragment charges $Z_{\rm{L}}$ and $Z_{\rm{L}}-1$, respectively, one obtains the relation
\begin{equation*}
C^{Z_{\rm{L}}}\left({\mathcal D}\right)-C^{Z_{\rm{L}}-1}\left({\mathcal D}\right)\cong \frac{Z_{\rm{f}}-2Z_{\rm{L}}+1}{Z_{\rm{L}}\left(Z_{\rm{f}}-Z_{\rm{L}}\right)}C^{Z_{\rm{L}}}({\mathcal D}).
\end{equation*}
From this relation, for the same shape of scission configuration, one can show that
\begin{equation*}
C^{Z_{\rm{L}}-1}\left({\mathcal D}\right)<C^{Z_{\rm{L}}}\left({\mathcal D}\right).
\end{equation*}
In consequence, if one assumes that 
\begin{equation*}
D^{Z_{\rm{L}}-1}\left({\mathcal D}\right)=D^{Z_{\rm{L}}}\left({\mathcal D}\right)
\end{equation*}
one can show that
\begin{equation*}
P^{Z_{\rm{L}}-1}\left({\mathcal D}\right)<P^{Z_{\rm{L}}}\left({\mathcal D}\right).
\end{equation*}
See Fig. \ref{fig3}. Actually, fragment deformation energy and Coulomb interaction energy between fragments are limited by the $Q$-value of the reaction.  The maximal Coulomb interaction energy corresponding to $Z_{\rm{L}}$ ($C^{Z_{\rm{L}}}_{\rm{max}})$  and the minimal value of deformation energy $D^{Z_{\rm{L}}}_{\rm{min}}$ obeys the relation
\begin{equation*}
C^{Z_{\rm{L}}}_{\rm{max}}=Q-D^{Z_{\rm{L}}}_{\rm{min}}.
\end{equation*}
Similarly, the relation corresponding to fragmentation with light fragment charge $Z_{\rm{L}}-1$ will be
\begin{equation*}
C^{Z_{\rm{L}}-1}_{max}=Q-D^{Z_{\rm{L}}-1}_{\rm{min}}.
\end{equation*}
Because $D$ increases with ${\mathcal D}$, one can show that
\begin{equation*}
D^{Z_{\rm{L}}-1}_{\rm{min}}<D^{Z_{\rm{L}}}_{\rm{min}}
\end{equation*}
then
\begin{equation*}
C^{Z_{\rm{L}}-1}_{\rm{max}}>C^{Z_{\rm{L}}}_{\rm{max}}.
\end{equation*}
See Fig. \ref{fig3}. Therefore, it is expected that among isobaric splits having similar $Q$- values, the more asymmetric charge  split will reach a more compact configuration, which corresponds to a lower deformation energy, a higher Coulomb interaction energy and, in consequence, a higher maximal total kinetic energy. \\\\
It was shown evidences that the even-even fragments couple (${}^{104}$Mo, ${}^{130}$Sn) reaches the $Q$-value in the reaction $^{233}$U($n_{\rm{th}}$,f). However, fragment charge was not measured. Nevertheless, the gap between the $Q$-value corresponding to (${}^{104}$Mo, ${}^{130}$Sn) and the other isobaric splits, respectively, is high enough to assume that the maximum value of the total kinetic energy is reached by this pair of fragments. \\\\
In 1988 U. Quade {\it et al.} studied cold fragmentation in the reaction $^{233}$U($n_{\rm{th}}$,f) \cite{9}. They measured the charge yield for isobaric splits. For the mass split ${\rm 89/145}$, between the two odd charge splits referred to $Z_{\rm{L}}=35\ $and $Z_{\rm{L}}=37$, respectively, although $Z_{\rm{L}}=35\ $corresponds to a $Q$-value $1$ MeV lower than the corresponding to $Z_{\rm{L}}=37$, its probability is higher in the coldest region. See Fig. \ref{fig4}.\\\\
Similarly, for the mass split ${\rm 94/140}$, between the two odd charge splits referred to $Z_{\rm{L}}=37$ and $Z_{\rm{L}}=39$, respectively, having a similar $Q$-value, in the cold fission region, the probability for $Z_{\rm{L}}=37$ is higher than the corresponding to $Z_{\rm{L}}=39$. See Fig. \ref{fig5}.\\\\
For the more asymmetrical mass split ${\rm 81/153}$, although $Z_{\rm{L}} =32$ corresponds to a $Q$-value approximately 2 MeV lower than the corresponding to $Z_{\rm{L}} =33$, its probability is higher in the coldest fission region. See Fig. \ref{fig6}.\\\\
Similarly, for the mass split $82/152$, although $Z_{\rm{L}} =32$ corresponds to a $Q$-value approximately 4 MeV lower than the corresponding to $Z_{\rm{L}} =34$, its probability is higher in the coldest region. Moreover, the slope of charge yield  as a function of kinetic energy corresponding to $Z_{\rm{L}}=32$ is higher than the corresponding to the higher charge $Z_{\rm{L}}=33$. See Fig. \ref{fig7}.\\\\
U. Quade {\it et al.} noticed, in cold fragmentations, the preferential formation of the element with the highest $Q$-value, as it is presented on Figs. \ref{fig4} and \ref{fig5}. However, among the 28 masses they found 10 exceptions. One can observe that the highest probability corresponds to a light fragment charge lower that the corresponding to the highest $Q$-value. In Figs. \ref{fig6} and \ref{fig7} are presented 2 of those 10 exceptions. In Tab. \ref{tbl:tab1} are showed their corresponding masses, the charges that maximize the isobaric $Q$-value, and the charge corresponding to the higher yield at the kinetic energy equal to 110.5 MeV. These results agree with the Coulomb hypothesis.\\\\
In 1994, W. Schwab {\it et al.} show that, for in the reaction $^{233}$U($n_{\rm{th}}$,f), definitely there is a clear trend to prefer more asymmetric charge split in cold fission \cite{11}.\\\\
J. Trochon {\it et al.} \cite{12, 13} measured $K_{\rm{max}}$($A_{\rm{L}}$) in the reaction $^{235}$U($n_{\rm{th}}$,f). They observed that, for isobaric fragmentation, the highest kinetic energy is reached by the charge corresponding to the highest $Q$-value; except for the mass 91 for which the charge that maximize $Q$ is 37, but the highest $K$ is reached by the charge 36 [12]. Moreover, they observed that only the odd charge split $51/41$ reach a ``true'' cold fission ($K_{\rm{max}}\cong Q$), while $K_{\rm{max}}$ referred to the magic charge split $50/42$  is ${\rm 3}$ MeV lower than the corresponding $Q$-value [13]. See Fig.~\ref{fig8}\\\\
\section{Conclusion}
It was shown that, in the cold region of thermal neutron induced fission of ${}^{233}$U and ${}^{235}$U, respectively, among isobaric charge fragmentations with similar $Q$-values of the reaction, the more asymmetric charge fragmentation will reach the higher maximal total kinetic energy. This results is interpreted, in a scission point model, as a ``Coulomb effect'' \cite{5, 6, 7}: a lower light fragment charge corresponds to a lower Coulomb repulsion, which will permit to reach a more compact configuration and, as a consequence, a lower minimal deformation energy, and a higher maximal Coulomb interaction energy. The result of that will be a higher maximal fragment kinetic energy. \\\\
%
%
% \begin{acknowledgments}
% %
%
% %
% \end{acknowledgments}
%\bibstyle{apsrev.bst}
%\bibliography{mylit}% Produces the bibliography via BibTeX.
%

\begin{table}
 \centering
 \caption{${}^{233}$U(n${}_{th}$,f). The ten light fragment masses for which the highest probabilities, at kinetic energy of 110.5 MeV, correspond to charges lower than the referred to the highest $Q$-value.  This table is based on information from \cite{9}}
\centering

\begin{tabular}{rrr}  
A & Z of highest  & Z of highest  \\
  & $Q$-value &  yield \\  
82 & 34 & 33 \\ 
86  & 36  & 34 \\ 
82  & 34  & 33 \\ 
87  & 36  & 35 \\ 
92  & 38  & 37 \\ 
102  & 42  & 40 \\ 
103  & 42  & 41 \\ 
85  & 34,35  & 34 \\  
81  & 33  & 32 \\ 
91  & 37  & 36 \\ 
101  & 41  & 40 \\ 
\label{tbl:tab1}
\end{tabular}
\end{table}
\end{document}